%% file: main.tex
\documentclass[runningheads]{llncs}
\usepackage[T1]{fontenc}
\usepackage{graphicx}
\usepackage{cite}
\usepackage{amsmath,amssymb,amsfonts}
\usepackage{algorithmic}
\usepackage{graphicx}
\usepackage{textcomp}
\usepackage{xcolor}
\usepackage{tikz-uml}
\usepackage{hyperref} 
\usepackage{lipsum}
\usepackage{comment}
\usepackage{listings}

\usepackage[export]{adjustbox}

\definecolor{codegreen}{rgb}{0,0.6,0}
\definecolor{codegray}{rgb}{0.5,0.5,0.5}
\definecolor{codepurple}{rgb}{0.58,0,0.82}
\definecolor{backcolour}{rgb}{0.95,0.95,0.92}

\lstdefinestyle{mystyle}{
    backgroundcolor=\color{backcolour},   
    commentstyle=\color{codegreen},
    keywordstyle=\color{magenta},
    numberstyle=\tiny\color{codegray},
    stringstyle=\color{codepurple},
    basicstyle=\linespread{1}\ttfamily\scriptsize,
    breakatwhitespace=false,         
    breaklines=true,                 
    captionpos=b,                    
    keepspaces=true,    
    morekeywords={def,for,while,return},
    emph={map,GPU_ThreadBlock,GPUdevice,ScheduleType,@ScheduleType,@ScheduleType.GPUdevice,MapCollapse,MapExpansion,MapFusion,InLocalStorage},
    emphstyle=\color{blue},     
    numbers=left,                    
    numbersep=5pt,
    showspaces=false,                
    showstringspaces=false,
    showtabs=false,                  
    tabsize=2,
    frame=single,
}    
    
\usepackage{microtype}

\def\BibTeX{{\rm B\kern-.05em{\sc i\kern-.025em b}\kern-.08em
    T\kern-.1667em\lower.7ex\hbox{E}\kern-.125emX}}
\begin{document}
\title{Portable High-Performance Kernel Generation for a Computational Fluid Dynamics Code with DaCe}
\titlerunning{Portable High-Performance CFD with DaCe}
%
\author{Måns~I.~Andersson\inst{1,3}\orcidID{0000-0002-6384-2630}\and
Martin~Karp\inst{1}\orcidID{0000-0003-3374-8093}\and Niclas~Jansson\inst{2}\orcidID{0000-0002-5020-1631}\and Stefano~Markidis\inst{1}\orcidID{0000-0003-0639-0639}}
\authorrunning{M. I. Andersson et al.}
%
\institute{KTH Royal Institute of Technology \and PDC Centre for High Performance Computing \and
\email{mansande@kth.se}}


%
\maketitle              
\begin{abstract}
With the emergence of new high-performance computing (HPC) accelerators, such as Nvidia and AMD GPUs, efficiently targeting diverse hardware architectures has become a major challenge for HPC application developers. The increasing hardware diversity in HPC systems often necessitates the development of architecture-specific code, hindering the sustainability of large-scale scientific applications. In this work, we leverage DaCe, a data-centric parallel programming framework, to automate the generation of high-performance kernels. DaCe enables automatic code generation for multicore processors and various accelerators, reducing the burden on developers who would otherwise need to rewrite code for each new architecture. Our study demonstrates DaCe's capabilities by applying its automatic code generation to a critical computational kernel used in Computational Fluid Dynamics (CFD). Specifically, we focus on Neko, a Fortran-based solver that employs the spectral-element method, which relies on small tensor operations. We detail the formulation of this computational kernel using DaCe's Stateful Dataflow Multigraph (SDFG) representation and discuss how this approach facilitates high-performance code generation. Additionally, we outline the workflow for seamlessly integrating DaCe's generated code into the Neko solver. Our results highlight the portability and performance of the generated code across multiple platforms, including Nvidia GH200, Nvidia A100, and AMD MI250X GPUs, with competitive performance results. By demonstrating the potential of automatic code generation, we emphasize the feasibility of using portable solutions to ensure the long-term sustainability of large-scale scientific applications. 
\end{abstract}

\section{Introduction}In the current high-performance computing landscape, the increased usage of accelerators, most importantly Graphics Processing Units (GPUs), has become a defining feature of modern supercomputers. The Roadrunner supercomputer at Los Alamos National Laboratory marked a significant milestone as the first system to shatter the petaflop performance barrier. Powered by 6,480 AMD Opteron dual-core processors and 12,960 IBM PowerXCell 8i accelerators, the Roadrunner supercomputer led to a new era in computational characterized by heterogeneous computing capabilities. With the advent of GPUs, Nvidia's CUDA programming framework emerged as a critical tool for optimizing scientific applications, yielding performance improvements. Today, we reach exascale capabilities with the Frontier supercomputer, featuring AMD MI250x GPUs. Yet, this transition has introduced a fresh challenge in the form of ROCm HIP, another programming interface required to harness the full power of these accelerators. Concurrently, other supercomputers have embraced Field-Programmable Gate Arrays (FPGAs), each demanding a distinct interface. Consequently, software developers grapple with the continuous need to adapt and rewrite code to suit the latest computing devices and their associated programming interfaces.

Addressing this challenge of new accelerators to be supported and exploited but HPC applications necessitates selecting and using efficient programming systems or approaches that allow the application developer to avoid rewriting the application. One possibility is to use compiler directives or code annotations that signal parallelization or porting requirements for accelerators. In this paradigm, the compiler generates the necessary code transformations. OpenMP represents an example of this approach. However, it comes with a significant performance limitation – the compiler's inability to optimize specific code segments due to its lack of algorithmic insight. This limitation serves as a driving force for the development and adoption of MLIR. 

An alternative strategy we employ in this research involves constructing a single-source, object-oriented framework for scientific applications. This framework encompasses a common interface for computational kernels that are amenable to diverse implementations targeting various hardware architectures, referred to as 'backends.' TensorFlow, a prominent exemplar of this approach, accommodates multiple backends tailored to specific accelerators. Our study extends this paradigm to the domain of Computational Fluid Dynamics, focusing on the Neko CFD code as an illustrative example \cite{jansson2024neko,karp2023large,10.1145/3581784.3627039,karp2024experience}. The drawback of this approach is that the programmer needs to write the code using multiple programming models and languages. The approach we use in this work is DaCe, a data-centric programming framework \cite{dace} that allows us first to express an algorithm as an SDFG, allowing us to decouple the algorithm formulation and performance optimization and porting to different architectures and second to generate automatically code for each system. The main objective of this work is to enable automatic efficient HPC code generation for a large-scale computational fluid dynamics code, Neko, and evaluate the effectiveness of this approach. The major contributions of this work are the following:
\begin{itemize}
\item We express and formulate the main Neko computational kernel, the matrix evaluation, as a DaCe SDFG and generate optimized code for different accelerator systems.
\item We design and develop a workflow to integrate DaCe code into the Neko CFD solver.
\item We evaluate the performance of the DaCe-generated code on different accelerated systems, including an Nvidia GH200, Nvidia A100, and AMD MI250x GPU systems and we compare them with the hand-crafted Neko kernels.
\end{itemize}
\section{Background}
In this section, we discuss the background material and previous related put in context our work on using automatic code generation and integrating into a large fluid dynamics code, called Neko.

\begin{figure*}[ht]
\centering
\resizebox{0.8\linewidth}{!}{\input{TikzFigs/DaCePipe}}
\caption{The SDFG for the Ax kernel (\texttt{ax.sdfg}) is generated from a restricted Python formulation. In general, the SDFGs can be generated and compiled Just-in-Time. But in this case, we generate the SDFG apply optimization transforms, and store the transformed SDFGs in a compressed format.}
\label{fig:dacecomp0}
\end{figure*}
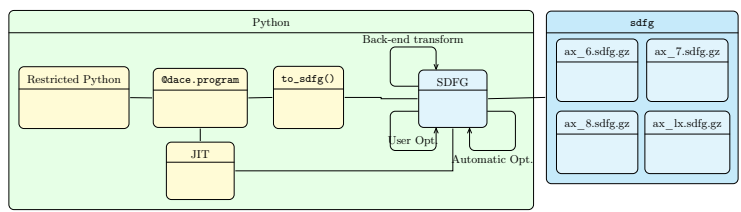
The Data-Centric Programming Framework, DaCe \cite{dace} is used to generate optimized code to be integrated into the CFD code. DaCe programming framework is actively developed at ETH Zürich and has been used in several applications~\cite{10.1145/3581784.3613214,andersson2023case,andersson2024towards}. The most famous application wholly developed in DaCe is the OMEN code for material science studies. The OMEN code was awarded the prestigious Gordon Bell Award for enabling full-machine Summit simulations.
The DaCe framework is based on an Intermediate Representation (IR) called SDFG which is designed to express data flow and portability of parallelism. The IR is generated from a high-level Python front-end and generates a graphical graph representation of the program. 
\begin{figure*}[h!]
\centering
\resizebox{1\linewidth}{!}{\input{TikzFigs/DaCeCompilation}}
\caption{The library is shipped with several ax.sdfg (one for each polynomial order) files. Additional optimization passes can be applied to ax.sdfg before compiling with \texttt{sdfgcc}. Configuration of the compilation is best controlled with DaCe environment variables. }
\label{fig:dacecomp}
\end{figure*}
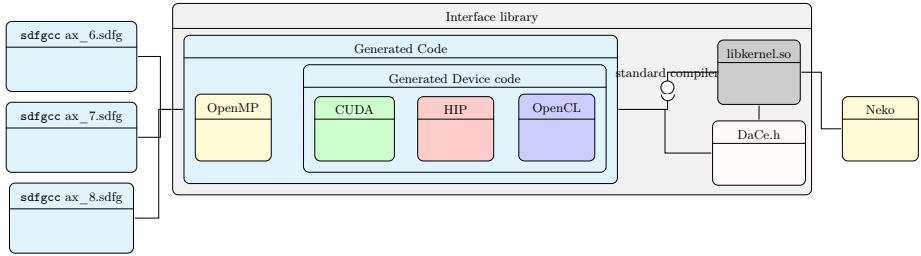
The generation of kernels on the IR form is found in~\autoref{fig:dacecomp0}. The SDFG has its own code generation and a supporting library for multiple hardware backends~\autoref{fig:dacecomp}.  
\subsection{Neko and Small Tensor Kernel} Neko is a new updated CFD code with roots from Nek5000. It exhibits excellent parallel performance for large-scale runs, including full-scale LUMI \cite{10.1145/3581784.3627039}. The code has been validated with several tests and applied to different grand challenges, including the Flettner rotor and Rayleigh-Bénard convection case. Differently from its ancestor, Neko code is written in modern Fortran, employing dynamic memory management and object-oriented structure exploiting one unified interface for implementations for different back ends. Neko is based on the spectral element method. In this method, the simulation is divided into several elements that are discretized with high-order basis functions. The two main ingredients of the Neko code are the matrix-multiply and the gather-scatter operation. The matrix-multiply is typically of small size as the size depends on the order of the spectral element.

\section{The Spectral Element Method} This text investigates the performance portability of a single test-kernel that arises 
in the CFD code Neko. We will investigate a reduced problem which is only a part of the full Navier-Stokes solver that Neko uses, namely the computational part that generates the matrix-free evaluation of the Poisson equation which is a major component. Here is the standard weak form formulation of Poisson's equation, which is the starting point for any Finite Element Method (FEM) solver. 
\begin{align}
    \int_{\Omega} \nabla u : \nabla v \operatorname{d\Omega} = \int_{\Omega} f v \operatorname{d\Omega}, \quad \forall v \in V,
\end{align}
where $\Omega$ is the domain and $v$ any test function in $V$.
The spectral element method is a FEM method based on a specific choice of elements that uses N\textsuperscript{th}-order Legendre polynomials $l_i$ interpolated on the Gauss-Lobatto-Legendre (GLL) quadrature points $\xi$ that lives on the reference element. 
\begin{align}
 \label{eq:li}
    l_i(\xi) = \frac{N(1-\xi^2 L_N(\xi) )}{(N+1)(\xi-\xi_i)L_N(\xi_i)}, \quad \xi \in [-1,1]. 
\end{align}
The spectral element solution $u_h$ that lives in $V_h$ can then be expressed as a tensor product on the reference element ($u_h^e$), 
\begin{align}
    u_h^e(\xi,\eta,\gamma) = \sum_{i,j,k}^N u_{ijk}l_i(\xi)l_j(\eta)l_k(\gamma).
\end{align}

\section{Methodology}
We compare with the kernels used in the latest release of Neko (v0.9.99). Which relies on hand-written code for each supported back end. The baseline implementations are the CUDA and HIP implementations used in Neko. Neko supplies two algorithms chosen at runtime with auto-tuning based on timing, this choice can be performed by setting \texttt{NEKO\_AUTOTUNE} to the preferred method (\texttt{1D} or \texttt{KSTEP}), the difference between the implementations are mainly related to how to best utilize the shared memory of the GPU. 

\subsection{Integration With Neko}
In \autoref{fig:dacecomp} we have a diagram to show the different steps from the generation of the SDFG to the compilation of the final library linked to Neko. Neko was designed with performance portability in mind -- that is, it was created with the intention that each hardware backend should be possible to be added with limited boilerplate code. In \autoref{lst:interface} we show how the C-to-Fortran interface is used in Neko, each backend's implementation has a wrapper to the corresponding execution call. As the focus is on accelerator offloading we annotate the Python code with \texttt{StorageType.GPU\_Global} to indicate that the generated SDFG's input parameters live on the device.

\begin{lstlisting}[language=Fortran,label=lst:interface, caption={The interface with a backend library using device pointers. In this case for the DaCe implementation of the kernel. The DaCe implementation does not require any boilerplate wrapping code for different GPU backends.}, float=h!]
interface
 subroutine dace_x_helm(handle, wd, ud, &
      dxd, dyd, dzd, dxtd, dytd, dztd, &
      h1d, g11d, g22d, g33d, g12d, g13d, g23d, nelv, lx) &
      bind(c, name='__dace_ax_helm')
   use, intrinsic :: iso_c_binding
   type(c_ptr), value :: handle
   type(c_ptr), value :: wd, ud
   type(c_ptr), value :: dxd, dyd, dzd
   type(c_ptr), value :: dxtd, dytd, dztd
   type(c_ptr), value :: h1d, g11d, g22d, g33d, g12d, g13d, g23d
   integer(c_int) :: nel, lx
 end subroutine dace_ax_helm
end interface
\end{lstlisting}

\subsection{The Kernel Expressed in DaCe}

DaCe is in principle meant to represent kernels as close to the mathematical formulation as possible and offloads optimizations to the compiler-like infrastructure of the intermediate representation SDFG. In order to work with the current code-base we have started from an existing implementation and simplified it. 
The re-worked implementation can be found in \autoref{lst:Ax}, note that the kernel is naively split into two parallel loops over all elements (\texttt{e1} and \texttt{e2})

\begin{lstlisting}[language=python, caption={The DaCe kernel written in the Python frontend, adapted from the standard kernel but simplified to easily be expressed with the DaCe formalism. Note the temporary arrays added on lines 12 to 17.},label={lst:Ax}, float=h!]
dtype = dace.float64
@dace.program()
def ax(wd : dtype[ne,lx,lx,lx],  ud : dtype[nel,lx,lx,lx], 
     dxd  : dtype[lx,lx],  				dyd  : dtype[lx,lx], 
     dzd  : dtype[lx,lx],         dxtd : dtype[lx,lx],
     dytd : dtype[lx,lx],         dztd : dtype[lx,lx],
     g11d : dtype[nel,lx,lx,lx],  g22d : dtype[nel,lx,lx,lx],
     g33d : dtype[nel,lx,lx,lx],  g12d : dtype[nel,lx,lx,lx], 
     g13d : dtype[nel,lx,lx,lx],  g23d : dtype[nel,lx,lx,lx])
     h1d  : dtype[nel,lx,lx,lx]): 
    
    # Temp arrays
    stmp   = np.empty((nel,lx,lx,lx),dtype=dtype) 
    rtmp   = np.empty((nel,lx,lx,lx),dtype=dtype)
    ttmp   = np.empty((nel,lx,lx,lx),dtype=dtype)
    urtmp  = np.empty((nel,lx,lx,lx),dtype=dtype) 
    ustmp  = np.empty((nel,lx,lx,lx),dtype=dtype) 
    uttmp  = np.empty((nel,lx,lx,lx),dtype=dtype)   

    # First map over all elements for each spatial dimension
    for e, k, j, i in dc.map[0:ne,0:lx,0:lx,0:lx] @ScheduleType.GPUdevice :  
        rtmp[e,k,j,i] = 0.0
        stmp[e,k,j,i] = 0.0
        ttmp[e,k,j,i] = 0.0

        # Sequential most inner loop
        for l in range(lx):
           rtmp[e,k,j,i] = rtmp[e,k,j,i]+dxd[l,i]*ud[e,k,j,l]
           stmp[e,k,j,i] = stmp[e,k,j,i]+dyd[l,j]*ud[e,k,l,i]
           ttmp[e,k,j,i] = ttmp[e,k,j,i]+dzd[l,k]*ud[e,l,j,i] 
        
        G00 = g11d[e,k,j,i]; G01 = g12d[e,k,j,i]; G02 = g13d[e,k,j,i]
        G11 = g22d[e,k,j,i]; G12 = g23d[e,k,j,i]; G22 = g33d[e,k,j,i]
        H   = h1d[e,k,j,I]
        rtmp_s = rtmp[e,k,j,I]
        stmp_s = stmp[e,k,j,I]
        ttmp_s = ttmp[e,k,j,i]

        urtmp[e,k,j,i] = H * (G00*rtmp_s + G01*stmp_s + G02*ttmp_s])
        ustmp[e,k,j,i] = H * (G01*rtmp_s + G11*stmp_s + G12*ttmp_s])
        uttmp[e,k,j,i] = H * (G02*rtmp_s + G12*stmp_s + G22*ttmp_s])

    # Second map over all elements for each spatial dimension
    for e2, k2, j2, i2 in dc.map[0:ne,0:lx,0:lx,0:lx]@ScheduleType.GPUdevice:
        wd[e2,k2,j2,i2] = 0.0
        for l2 in range(lx):
           wd[e2,k2,j2,i2] = wd[e2,k2,j2,i2] + dxtd[l2,i2]*urtmp[e2,k2,j2,l2]
           wd[e2,k2,j2,i2] = wd[e2,k2,j2,i2] + dytd[l2,j2]*ustmp[e2,k2,l2,i2] 
           wd[e2,k2,j2,i2] = wd[e2,k2,j2,i2] + dztd[l2,k2]*uttmp[e2,l2,j2,i2]           
\end{lstlisting}
The SDFG IR is based on a set of primitives, that are briefly described in this section. There is a human-readable visual representation of the SDFG IR that will mostly be omitted in this paper, the visual representation of the primitive is described in parentheses. The fundamental primitive is the state (square/rectangle) which is connected. Data movement is handled by the \texttt{Memlet} (arrow) which indicates the movement of data between containers denoted with an arrow in the \texttt{sdfg}. Parallelism is indicated by \texttt{Map} (split hexagon), where in the simplest case everything inside a map can be executed embarrassingly parallel or with a \texttt{Write-Conflict Resolution} (dotted line). The DaCe programming model often leads to formulations that require transient \texttt{data containers} (elipse) to be removed in an optimization step. The \texttt{Tasklets} (hexagon) are the smallest units of the \texttt{sdfg} and are not shown in the \autoref{fig:sdfg_tranformation} as it is deemed too granular to be of interest for this study.

\subsection{Optimization With SDFG}
\begin{figure*}[t!]
    \centering
    \includegraphics[width=1\linewidth,frame]{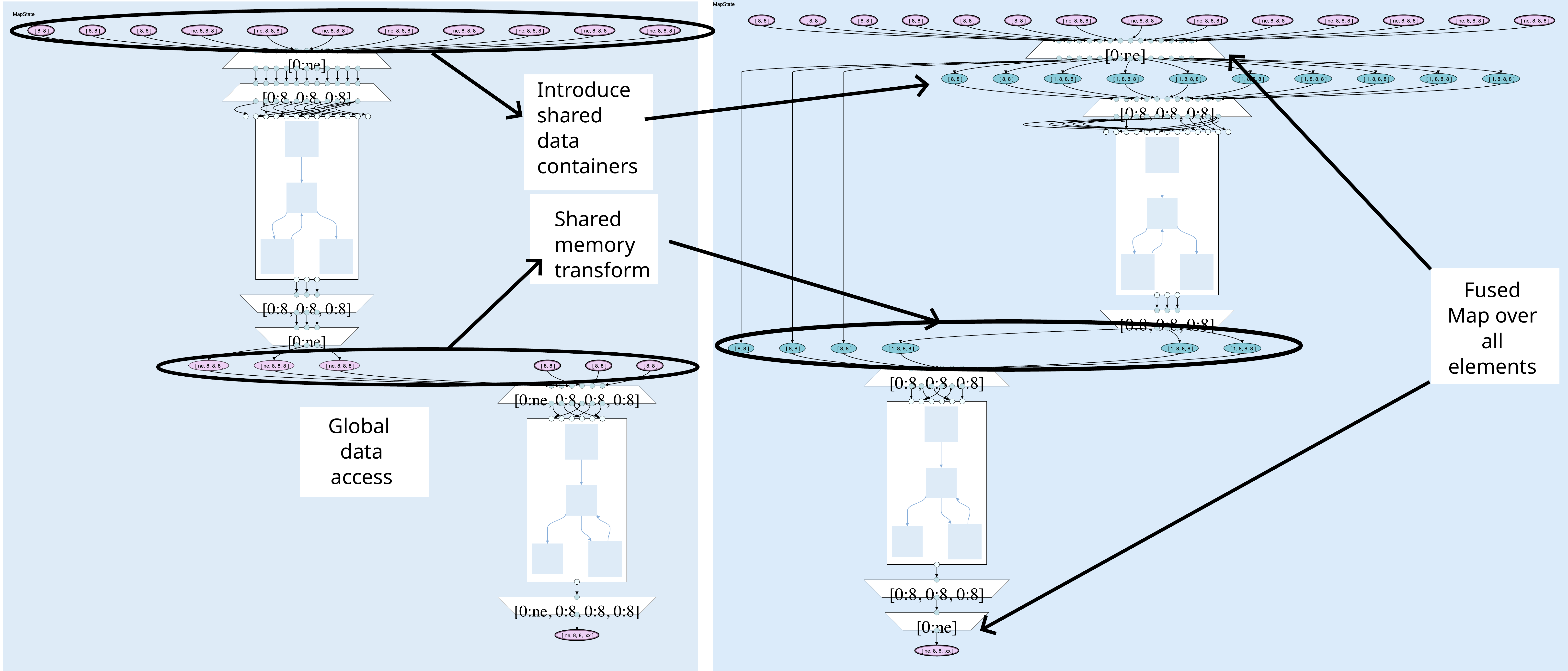} 
    \caption{An example of the human-readable IR transformed from the naive SDFG implementation to the 3D parallelization. A data container is eplise-shaped. The \textcolor{cyan}{dark cyan} represents shared memory data containers and \textcolor{magenta}{dark pink} represents global GPU data containers. The promotions of memory are allowed since all data movement occurs within the same parallel map. }
    \label{fig:sdfg_tranformation}
\end{figure*}

The parallelization strategy that is described in the coming section is a result of using a greedy approach to GPU optimization of SDFGs, merging maps when possible, and introducing shared memory containers when possible. 
\begin{table}[ht!]
\caption{A brief summary to the transforms used in the coming section.}
    \label{tab:opt}
    \centering
    \begin{scriptsize}
\begin{tabular}{p{0.25\linewidth}  p{0.55\linewidth}}
 & \\
{Map transformations} & \\ \hline 
\texttt{MapFusion} & Fuses two consecutive maps, if the maps align.\\ 
\texttt{MapCollapse} & Collapses two nested maps into one. The new map has the union of the dimensions of the original maps\\
\texttt{MapExpansion} & Expands a map into the different iterators and enables a hierarchical view of parallelism.\\
\texttt{MapTiling}  &  Implements the orthogonal tiling, which is a type of nested map fission that creates tiles in every dimension of the matched map.\\
\texttt{StripMining} & Allows loop-fracturing which can allow vectorization and improved cache usage with loop-blocking.\\
\texttt{WarpTiling} & A specific transform that combines StripMining and MapTiling. \\
{Data transformations } & \\ \hline
\texttt{LocalStorage} & Introduces a local transient array for caching data\\
\texttt{}
& \\
{Flow transformations} & \\  \hline
\texttt{StateFusion} & Fuses two states into a single state. \\
\texttt{MapToForLoop} & Converts map to for-loop. 
\end{tabular}
\end{scriptsize}
\end{table}
The data-centric programming model provided by DaCe forces the programmer to write the kernel in a general way first, premature optimizations in the creation of the SDFG often hinder the ability to improve the performance by SDFG transforms. Some of the most common transformations are summarized in \autoref{tab:opt}. This work focuses on three aspects: 1) fusing the two \texttt{maps} over all elements 2) \texttt{block~tiling} along the spatial dimensions \textit{i}, \textit{j}, and \textit{k}, and 3) shared memory transform to remove the transient data containers that were introduced when converting the kernel DaCe. In \autoref{fig:sdfg_tranformation} we present the SDFG in the human-readable format before and after the passes described in \autoref{lst:pass}, the notable shared memory transform is highlighted with color.
\begin{lstlisting}[float=h!,language=Python,caption={The first function treats the first part of the code. It has two main tasks: to promote data containers to shared memory and to prepare for the fusion of the \texttt{e1} and \texttt{e2}. The second function treats the remaining code. It has two main tasks: to promote data containers to shared memory and to fuse the \texttt{e1} and \texttt{e2} maps.},label={lst:pass}]
def ax_3D_optimization_1(sdfg : dc.SDFG, lx_val : int)    
    sdfg.apply_gpu_transformations()
    sdfg.apply_transformations(MapExpansion) # Expand Maps 'e','i','j','k'
    sdfg.apply_transformations(MapCollapse)  # Recollapse Maps 'i' and 'j'
    sdfg.apply_transformations(MapCollapse)  # Recollapse Maps 'i,j' and 'k'
    sdfg.replace('lx',str(lx_val))           # Specify symbol to a constant
                                             # enables constant propagation 
    entry = find_map_by_param(sdfg, 'e') 
    exit  = find_map_by_param(sdfg, 'k')
    exit.schedule = dc.ScheduleType.GPU_ThreadBlock #Promotes thread-level  
    
    for arr in ['ud', '', ..., '']:
       InLocalStorage.apply_to(sdfg,dict(array=arr),node_a=entry,node_b=exit)
       
def ax_3D_optimization_2(sdfg : dc.SDFG, lx_val : int)    
    entry = find_map_by_param(sdfg, 'e2')
    MapExpansion.apply_to(sdfg, map_entry=entry) 
    # Explicit Map collapse
    MapCollapse.apply_to(sdfg, outer_map_entry=find_map_by_param(sdfg, 'k2'),
                               inner_map_entry=find_map_by_param(sdfg, 'j2')) 
    MapCollapse.apply_to(sdfg, outer_map_entry=find_map_by_param(sdfg, 'j2'),
                               inner_map_entry=find_map_by_param(sdfg, 'i2')) 
    exit = find_map_by_param(sdfg,'k2')
    exit.schedule = dc.ScheduleType.GPU_ThreadBlock

    for arr in ['dxtd', '', ..]
       InLocalStorage.apply_to(sdfg,dict(array=arr),node_a=entry,node_b=exit)
    
    # Fuse the two outer Maps (over elements)  
    sdfg.apply_transformations(MapFusion) 
    sdfg.simplify()               
\end{lstlisting}

\subsection{Experimental Setup}
This study relies on three systems with Nvidia and AMD GPUs. We have two Nvidia nodes (NJ and Sleipner) in a local cluster which has previously been used to evaluate the performance of Nvidia hardware A100 \cite{10.1145/3468044.3468053}, and GH200 \cite{10.1145/3673038.3673110}. The AMD nodes are from the LUMI-G portion of the LUMI supercomputer at CSC’s data center in Kajaani, Finland. The hardware and software setups are summarized in \autoref{tab:setup} only one GCD is considered. The configuration of Neko for these experiments is straightforward as there are only a few dependencies required. Neko and the DaCe kernels are compiled with \texttt{double}.
\begin{table}[h!]
    \centering
    \caption{The test systems and software setup.} 
    \begin{tabular}{llllll} \hline 
       \textbf{System name} & \textbf{GPU} & \textbf{CUDA/ROCm} & \textbf{Compiler} & \textbf{DaCe} & \textbf{Neko} \\\hline 
       NJ     & A100 &  CUDA 11.5 & GCC 12 & 0.15.1$^{*}$ & 0.9.99  \\
       Sleipner& GH200 & CUDA 12.5 & GCC 13 & 0.15.1$^{*}$ & 0.9.99 \\
       LUMI-G & MI250X & ROCm 6.0.3 & Cray 16 & 1.0.1$^{**}$ & 0.9.99 \\\hline 
       \multicolumn{6}{l}{\parbox{.8\textwidth}{\tiny * We saw significant performance degradation for recent DaCe versions (0.16 and 1.0.1) on the two NVIDIA systems.}} \\
       \multicolumn{6}{l}{\parbox{.8\textwidth}{\tiny ** The older DaCe versions 0.15.1 and 0.16 did not generate the correct code or correct build configuration on the MI250X machine.}}
    \end{tabular}
    \label{tab:setup}
\end{table}
We perform evaluations on nine different-sized cubical meshes with 128, 256, 512, 1024, 2048, 4096, 8192, 16384, and 32768 elements ($N_E$) and $N_E(lx-1)^3$ unknowns per problem. We perform performance measurements for $3 \leq lx \leq 8$.
\section{Results} 
This study evaluates the three implementations with respect to performance measured in Gflops/s on the three different types of GPUs for the nine mesh sizes and six polynomial orders of basis functions.\\
\\
\begin{figure}[h!]
    \centering\includegraphics[width=0.99\textwidth,frame]{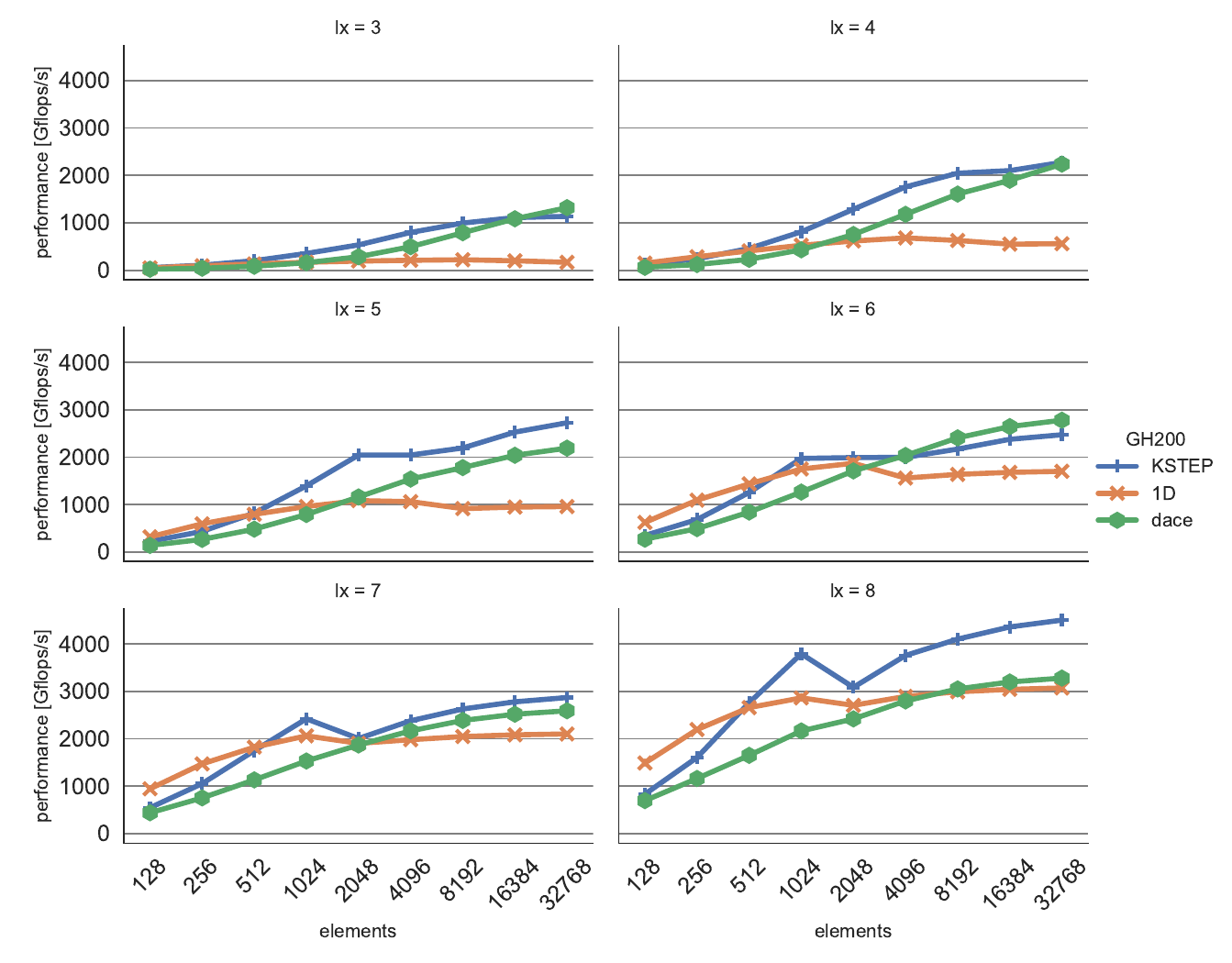}
    \caption{The performance (Gflops/s) for different mesh sizes and $3 \leq lx \leq 8$ with the different implementations on a GH200 node.}
    \label{fig:GH200}
\end{figure}

\noindent \textbf{GH200 (\autoref{fig:GH200}):} The parallelization strategy of the DaCe implementation performs better than the \texttt{1D} strategy for most of the mesh sizes above 2048 elements.  The \texttt{KSTEP} strategy performs best for most configurations although $lx=6$ is notable as DaCe is the highest-performing strategy for the large mesh sizes and the highest achieved performance. It is the \texttt{KSTEP} strategy that has the highest achieved performance, this occurs for the largest mesh size and $lx = 8$ which is expected. We note that the \texttt{KSTEP} has a distinct performance peak that is achieved for smaller meshes with higher $lx$ that is followed by a sharp decrease in performance with respect to the number of elements followed by a slow increase.\\
\\
\begin{figure}[t!]
    \centering\includegraphics[width=0.99\textwidth,frame]{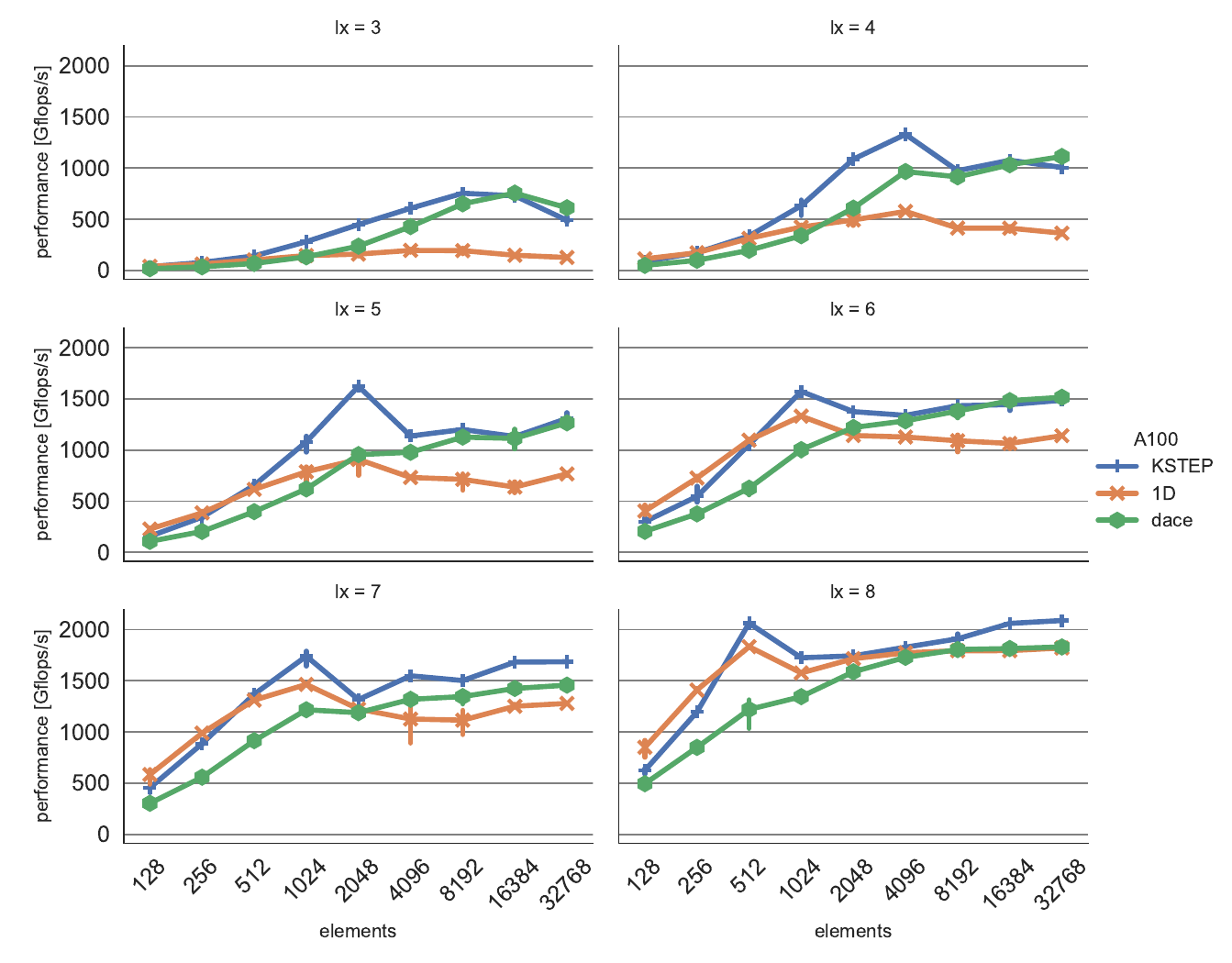}
    \caption{The performance (Gflops/s) for different mesh sizes and $3 \leq lx \leq 8$ with the different implementations on an A100 node.}
    \label{fig:A100}
\end{figure}

\noindent \textbf{A100 (\autoref{fig:A100}):} The results from the A100 machine have a larger standard deviation due to more background processes as it is a shared node. The general trends are similar to the GH200; however, DaCe and \texttt{KSTEP} perform more similarly for $lx < 7$. For the higher $lx$: \texttt{KSTEP} performs better. The \texttt{1D} implementation has consistently the lowest performance for large mesh sizes but also the best performance for very few elements. As opposed to the GH200 results the peak performance is often found at a small mesh size except for $lx = 3$. As opposed to the GH200 case the slow increase in performance after the dip never catches up to the peak performance\\

\begin{figure}[h!]
    \centering\includegraphics[width=0.99\textwidth,frame]{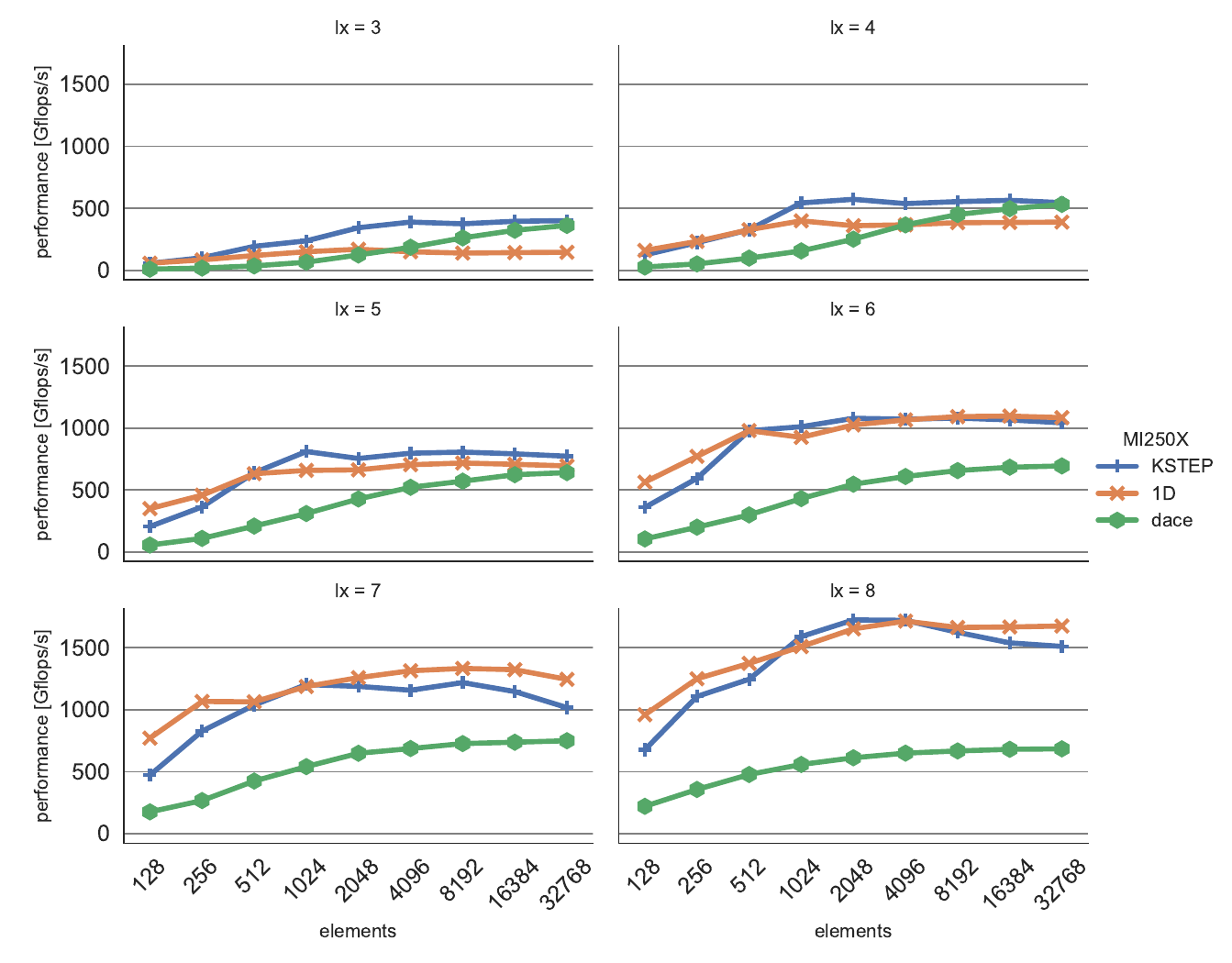}
    \caption{The performance (Gflops/s) for different mesh sizes  and $3 \leq lx \leq 8$ with the different implementations on an MI250X node using one GCD.}
    \label{fig:MI250X}
\end{figure}
\noindent \textbf{MI250X (\autoref{fig:MI250X}):} We note that the DaCe performance on the MI250X is significantly lacking compared with the base implementation, but, notably, all strategies perform poorly compared with the A100 results for peak performance. For $lx>6$ the \texttt{1D} strategy performs the best for most mesh sizes, this is in stark contrast to the Nvidia results. For small $lx$ the behavior is similar to the previous tests. The peak performance is a tie between \texttt{KSTEP} and \texttt{1D} for a medium-sized mesh and $lx=8$\\

\section{Related works} 
\noindent \textbf{Performance portability frameworks} 
As HPC systems continue to evolve with diverse hardware architectures, achieving performance portability across different processing devices has become increasingly important. Various approaches have been developed to tackle this challenge. Kokkos, a widely adopted C++ library, offers an abstraction layer for parallel execution and memory management across multiple hardware platforms, including CPUs and GPUs~\cite{trott2021kokkos}. Similarly, RAJA is a parallel programming framework designed to provide abstractions that enable performance portability across heterogeneous hardware~\cite{beckingsale2019raja}. Additionally, \texttt{alpaka}~\cite{alpaka2016} supports multiple execution models, such as CUDA, OpenMP, and SYCL. \textbf{Portable high-performance spectral element kernels and matrix free-kernels} 
The spectral elements code NekRS~\cite{ChalmersKarakusAustinSwirydowiczWarburton2020} relies on OCCA kernels \cite{abdelfattah2021gpu,fischer2020scalability} for portability (CUDA, HIP, OpenCL). Similar benchmark codes based on \texttt{Nekbone} has implementations for FPGA~\cite{9460528} and OpenCL~\cite{10.1145/3492805.3492818}. Low-order stencil-based matrix-free operator was studied in the portable \texttt{alpaka} based Poisson solver~\cite{pennati2025parallelhighlyportablehpcpoisson}. 

\section{Discussion and Conclusion}

We have shown an implementation of the main computational kernel of Neko's matrix evaluation in DaCe with a set of optimization passes that generates code that is competitive with some of the hand-tuned backend-specific implementations. We have integrated the generated code into a large-scale solver based on modern Fortran. We acknowledge that further optimizations can be done in line with previous literature, such as other parallelization strategies, for example, 2D or 1D parallelization strategies which are better suited for higher than $lx$ of order eight. We show successful runs on AMD systems, which unfortunately leaves some performance on the table. This might be due to the changes in DaCe between version 0.15.1 and 0.16 versions which degraded performance for the Nvidia version compared with the Neko kernels to a similar degree as we see on the AMD system. To conclude the performance is comparable with the highly tuned Neko kernels with a single source implementation that is easily tuneable. 

\begin{credits}
\subsubsection{\ackname} {\small The computations of this work were enabled by resources provided by the National Academic Infrastructure for Supercomputing in Sweden projects NAISS 2024/22-531 and NAISS 2024/9-20 and by the ScaLab Group for providing access to the Sleipner cluster. Financial support was provided by the Swedish e-Science Research Centre (SeRC) Exascale Simulation Software Initiative (SESSI). }
\end{credits}

%
%
%

\bibliographystyle{splncs04}
\bibliography{ref}
\end{document}

%% file: TikzFigs/DaCePipe.tex
\begin{tikzpicture}
\begin{umlstate}[name=Python, fill=green!10]{Python}
    \umlbasicstate[x = -4.5, y = 2.5 ,fill=yellow!20, name=RePython]{Restricted Python}{}    
    \umlbasicstate[x = -1,  y = 2.5 ,fill=yellow!20, name=DaCe]{\texttt{@dace.program}}{}    
    \umlbasicstate[x = -1,  y = 0.5,fill=yellow!20, name=JIT]{JIT}{}

    \umlbasicstate[x = 2,  y = 2.5,fill=yellow!20, name=tosdfg]{\texttt{to\_sdfg()}}{}
    \umlbasicstate[x = 6,  y = 2.5 ,fill=cyan!10, name=SDFG]{SDFG}{}

    \umltrans[recursive=-160|-120|1.9cm, recursive direction=left to bottom, arg={User Opt.}, pos=2.5]{SDFG}{SDFG}
    \umltrans[recursive=160|120|1.9cm, recursive direction=right to bottom, arg={Back-end transform}, pos=2.5]{SDFG}{SDFG}
    \umltrans[recursive=-20|-60|1.9cm, recursive direction=right to bottom, arg={Automatic Opt.}, pos=2.5]{SDFG}{SDFG}

    \umlassemblyconnector[interface=]{RePython}{DaCe}{}
    \umlassemblyconnector[interface=]{DaCe}{tosdfg}{}
    \umlVHVassemblyconnector[interface=]{DaCe}{JIT}{}

    \umlHVHassemblyconnector[interface=]{tosdfg}{SDFG}{}
    \umlHVassemblyconnector[interface=]{JIT}{SDFG}{}

\end{umlstate}
\begin{umlstate}[name=SDFGL, fill=cyan!20]{\texttt{sdfg}}
\umlbasicstate[x = 10,  y = 1.25,fill=cyan!10, name=ax7sdfg]{{ax\_8.sdfg.gz}}{}
\umlbasicstate[x = 10,  y = 3.25,fill=cyan!10, name=ax8sdfg]{{ax\_6.sdfg.gz}}{}
\umlbasicstate[x = 12.5,  y = 1.25,fill=cyan!10, name=ax8sdfg]{{ax\_lx.sdfg.gz}}{}
\umlbasicstate[x = 12.5,  y = 3.25,fill=cyan!10, name=ax8sdfg]{{ax\_7.sdfg.gz}}{}
\end{umlstate}
\umlassemblyconnector[interface=]{SDFG}{SDFGL}{}

\end{tikzpicture}

%% file: TikzFigs/DaCeCompilation.tex
\begin{tikzpicture}
\umlbasicstate[x = 12,  y = 0.25 ,fill=cyan!10, name=SDFGCompile]{\texttt{sdfgcc} ax\_8.sdfg}{}   
\umlbasicstate[x = 12,  y = 2.25 ,fill=cyan!10, name=SDFGCompile1]{\texttt{sdfgcc} ax\_7.sdfg}{}   
\umlbasicstate[x = 12,  y = 4.25 ,fill=cyan!10, name=SDFGCompile2	]{\texttt{sdfgcc} ax\_6.sdfg}{}   

\begin{umlstate}[name=KERlib, fill=gray!10]{Interface library}
    \begin{umlstate}[name=GENE, fill=cyan!10]{Generated Code}
        \umlbasicstate[x = 16, y=2.5, name=OMP, fill=yellow!20]{OpenMP}
        \begin{umlstate}[name=Device, fill=cyan!10]{Generated Device code}
            \umlbasicstate[x = 19, y=2.5, name=CUDA, fill=green!20]{CUDA}
            \umlbasicstate[x = 21.5, y=2.5, name=HIP, fill=red!20]{HIP}
            \umlbasicstate[x = 24 , y=2.5, name=OCL, fill=blue!20]{OpenCL}
        \end{umlstate}
        
    \end{umlstate}
\umlbasicstate[x = 29, y = 3.9, name=lib, fill=black!20]{libkernel.so}{}

\end{umlstate}
\begin{umlstate}[x = 29, y = 1.9, name=DACElib, fill=pink!10]{DaCe.h}
\end{umlstate}

\umlbasicstate[x=32, y=2.5]{Neko}{}

\umlVHVassemblyconnector[]{DACElib}{lib}{}{}
\umlHVHassemblyconnector[]{DACElib}{GENE}{}{}
\umlHVHassemblyconnector[interface=]{SDFGCompile}{GENE}{}{}
\umlHVHassemblyconnector[interface=]{SDFGCompile1}{GENE}{}{}
\umlHVHassemblyconnector[interface=]{SDFGCompile2}{GENE}{}{}

\umlHVHassemblyconnector[interface=standard compiler]{GENE}{lib}{}{}

\umlHVHassemblyconnector[]{lib}{Neko}{}{}



\end{tikzpicture}